\documentclass[acmlarge]{acmart}
\usepackage{colortbl}
\usepackage{makecell}
\usepackage{soul}
\newcommand{\rv}[1]{{\leavevmode\color{black}#1}}
\AtBeginDocument{%
  \providecommand\BibTeX{{%
    \normalfont B\kern-0.5em{\scshape i\kern-0.25em b}\kern-0.8em\TeX}}}

\setcopyright{acmcopyright}
\acmJournal{IMWUT}
\acmYear{2022}
\acmVolume{6} 
\acmNumber{3} 
\acmArticle{121} 
\acmArticleSeq{5}
\acmMonth{9} 
\acmPrice{15.00}
\acmDOI{10.1145/3550321}

\begin{document}

\title{Synapse: Interactive Guidance by Demonstration with Trial-and-Error Support for Older Adults to Use Smartphone Apps}
\thispagestyle{empty}

\def\markup{0}  
\if\markup 1
\usepackage{soul}
\newcommand{\rv}[1]{{\leavevmode\color{blue}#1}}
\newcommand{\camera}[1]{{\leavevmode\color{blue}#1}}
\else
\renewcommand{\rv}[1]{#1}
\renewcommand{\st}[1]{#1}
\newcommand{\camera}[1]{#1}
\fi

\author{Xiaofu Jin}
\orcid{0000-0002-7239-3769}
\affiliation{%
  \institution{The Hong Kong University of Science and Technology (HKUST)}
  \country{China}
}
\email{xjinao@connect.ust.hk}

\author{Xiaozhu Hu}
\orcid{0000-0003-3832-3713}
\affiliation{%
  \institution{The Hong Kong University of Science and Technology}
  \country{China}
}
\email{huxz19@tsinghua.org.cn}

\author{Xiaoying Wei}
\orcid{0000-0003-3837-2638}
\affiliation{%
  \institution{The Hong Kong University of Science and Technology}
  \country{China}
}
\email{xweias@connect.ust.hk}

\author{Mingming Fan}
\authornote{Corresponding Author}
\orcid{0000-0002-0356-4712}
\affiliation{
  \institution{The Hong Kong University of Science and Technology (Guangzhou)}
  \country{China}
}
\affiliation{
  \institution{The Hong Kong University of Science and Technology}
  \country{China}
}
\email{mingmingfan@ust.hk}

\renewcommand{\shortauthors}{Jin et al.}

\begin{abstract}
  
As smartphones are widely adopted, mobile applications (apps) are emerging to provide critical services such as food delivery and telemedicine. While bring convenience to everyday life, this trend may create barriers for older adults who tend to be less tech-savvy than young people. In-person or screen sharing support is helpful but limited by the help-givers' availability. Video tutorials can be useful but require users to switch contexts between watching the tutorial and performing the corresponding actions in the app, which is cumbersome to do on a mobile phone. Although interactive tutorials have been shown to be promising, none was designed for older adults. Furthermore, the trial-and-error approach has been shown to be beneficial for older adults, but they often lack support to use the approach. Inspired by both interactive tutorials and trial-and-error approach, we designed an app-independent mobile service, \textit{Synapse}, for help-givers to create a multimodal interactive tutorial on a smartphone and for help-receivers (e.g., older adults) to receive interactive guidance with trial-and-error support when they work on the same task. We conducted a user study with 18 older adults who were 60 and over. Our quantitative and qualitative results show that Synapse provided better support than the traditional video approach and enabled participants to feel more confident and motivated. Lastly, we present further design considerations to better support older adults with trial-and-error on smartphones.

\end{abstract}




  \begin{teaserfigure}
    \includegraphics[width=\textwidth]{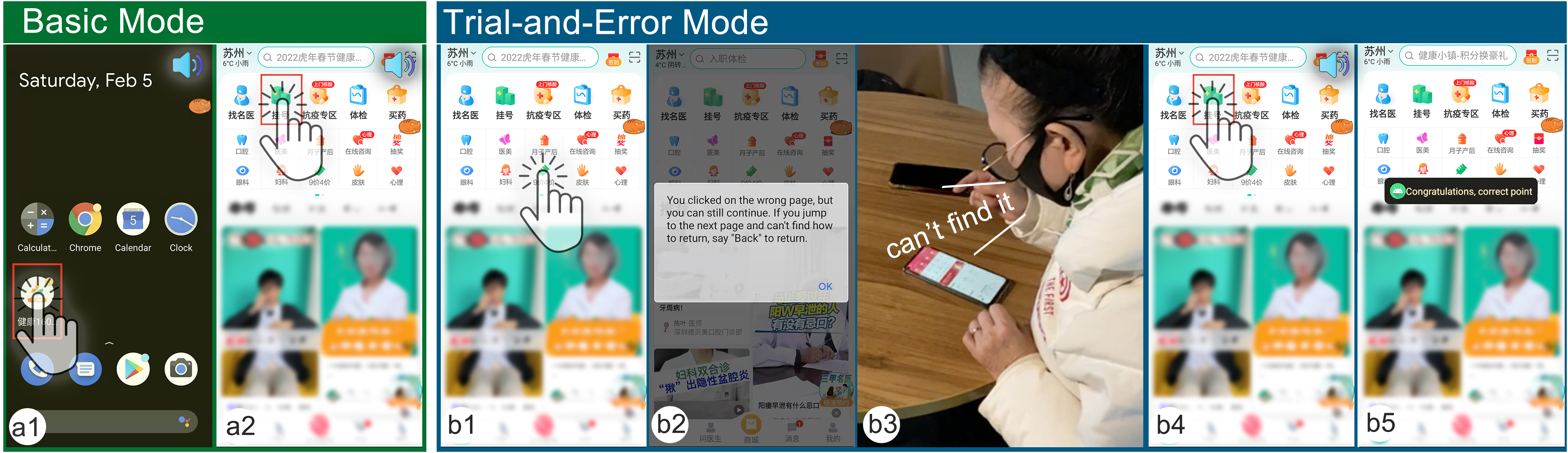}
    \caption{\textbf{Synapse} runs on a smartphone as a background service and helps older adults complete tasks by providing step-by-step interactive guidance (Basic mode) and trial-and-error support (Trial-and-Error mode), which are created by the demonstration of a help-giver. 
    In the Basic mode (a1, a2), Synapse highlights the target item with a red rectangle and plays the corresponding audio instructions recorded by a help-giver during the demonstration phase. 
    After the user clicks the highlighted icon (a1), the app proceeds to the next screen, where Synapse identifies and highlights the next target item (a2). In the Trial-and-Error mode, Synapse allows the user to freely explore the app. If they clicks a wrong item (b1), Synapse alerts them but still allows them to continue their exploration (b2). If they still could not complete the task or realized they made a mistake after a few trials, they could ask Synapse for help by saying, such as ``can't find it'' (b3). Synapse then returns to the last correct step with the corresponding audio instructions (b4). After they has completed the step correctly, Synapse provides positive multimodal feedback with both a textual prompt and an audio tone (b5).}
    \Description{figure description}
    \label{fig:teaser}
  \end{teaserfigure}

\maketitle

\section{Introduction}

Smartphones are widely adopted by people of all ages, and many critical services, such as food delivery and telemedicine, are increasingly offered through mobile applications (apps), in particular during the current COVID-19 pandemic. While making lives more convenient for many tech-savvy people, this trend may create barriers for older adults, who tend to be less tech-savvy than young people, to access these critical services via their smartphones timely and effectively. Indeed, many older adults were found to use smartphones only as feature phones~\cite{BerenguerAnabela2017ASUA} (e.g., making phone calls, sending text messages) and often need external support to overcome difficulties~\cite{Pang2015OAADS}. Although in-person support offered by help-givers (e.g., children or relatives) is helpful, it is limited by their availability. Furthermore, more and more older adults live independently, making it harder for them to get timely in-person support~\cite{UN17Arrangement}. Thus, to help older adults better overcome difficulties when using smartphones, other forms of support are needed. 

Video tutorials are common for people who are learning to use mobile apps, and researchers have investigated ways to lower the efforts to create better video tutorials (e.g., \cite{Pongnumkul2011,Chi2012Mix}). While following a video tutorial on a smartphone, users have to switch between watching the video and completing the corresponding task in the app, which requires them to remember the steps in the video in order to perform them in the app. This can be challenging for older adults, who often have declining physical and cognitive abilities. 
To reduce context switching costs, tools like Sketch-sketch Revolution ~\cite{Jennifer2011Sketch-Sketch} and DemoWiz ~\cite{Chi2014DemoWizRS} offer in-app interactive tutorials or within-context features. However, such approaches are app-dependent and not designed to support older adults in general learning to use mobile apps. 
Alternatively, a phone call with video sharing and remote control enabled by third-party apps (e.g., InkWire~\cite{InkWire}, TeamViewer~\cite{TeamViewer}, or AirDroid~\cite{AirDroid}) provides a general approach for the help-giver to view the \rv{help-receiver's} smartphone and take control of it remotely if needed. However, such methods require them to be available at the same time and thus are limited by the availability of \rv{the} help-giver. Furthermore, granting remote access to smartphones also raises privacy and security concerns~\cite{Vyas2019AppAmigo}. 

Compared to the above approaches, interactive tutorials created by demonstration (e.g., EverTutor~\cite{Wang2014EverTutor}, Remo~\cite{Remo}) offer step-by-step guidance within the app and can be accessed at any time. While Remo was a conceptual model without implementation, EverTutor provided a computer-vision-based approach to assist help-givers with creating interactive tutorials. However, none of these works focused on older adults. We extended this line of work and designed \textbf{Synapse} by focusing on supporting \rv{older adults as} help-receivers and allowing them to better leverage step-by-step interactive guidance. Unlike prior work, Synapse caters to older adults' needs. As older adults tend to have difficulty understanding technical terms~\cite{10.1145/3209882,bruder2006improving}, Synapse allows the help-giver to record plain-language audio instructions and attach them to the \rv{step-by-sep} interactive guide. Moreover, Synapse provides multi-modal feedback, such as visual highlights, textual instructions, and audio alerts, and corrects misclicks to cater to older adults' varied vision, hearing, and motor abilities. 

Furthermore, while trial-and-error is beneficial for \rv{learning} technology, older adults tend to be afraid of performing trial-and-error due to apprehension of making mistakes~\cite{barnard2013learning} and lack of error-recovery support~\cite{fan2018guidelines}. Synapse is the first attempt to support older adults in performing trial-and-error, including recognizing their errors and helping them recover from \rv{the} errors. \rv{By} inferring where older adults start to deviate from the demonstrated interaction path, Synapse constructs the error-recovery path and allows them to use natural language to ask for help (e.g., ``can't find it''). Additionally, Synapse actively provides feedback \rv{about} their operations, alerting them when they clicked on a wrong item and allowing for recovery with natural language (e.g., ``back'', ``start over''). 

To understand whether and to what extent Synapse's interactive guidance and trial-and-error support could help older adults complete tasks on smartphones, we conducted a user study with 18 older adults who are 60 and over to perform tasks with three conditions: traditional video tutorial, Synapse (Basic mode), and Synapse (Trial-and-Error mode). \rv{While the Basic mode provides interactive guidance with multi-modal feedback and voice assistance, the Trial-and-Error mode also allows for free exploration and provides error-recovery support.} Our results show that older adults completed tasks faster with both modes of Synapse compared to with the video tutorial. They found both interactive support (e.g., being able to ask questions in natural language, and multi-modal feedback) and trial-and-error support \rv{were} helpful and enjoyable. Moreover, they were more confident in exploring the app with trial-and-error support. We analyzed the errors they made and provided design considerations for further improving trial-and-error support.  
In sum, we make the following contributions:
\begin{itemize}

\item We designed a novel interactive app-independent service, \textit{Synapse}, that caters to older adults' needs by providing interactive step-by-step instructions \rv{as an overlayer} with multi-modal feedback, voice assistant, and trial-and-error support to assist them with completing tasks on smartphones.

\item We conducted a user study with older adults to understand their performance and user experience of the two modes of Synapse (i.e., Basic and Trail-and-Error) compared to video tutorials. Moreover, we identified common types of errors older adults made when using the trial-and-error approach and present design considerations for further improvement. 

\end{itemize}

\section{Related Work}
\subsection{Challenges, Learning needs, and Learning Preferences of Using Smartphones Among Older Adults}

\subsubsection{Challenges.}
Literature has revealed typical challenges faced by older adults in learning to use new technologies.
Echt et al. found that \textit{cognitive} declines in spatial working memory, as well as perceptual speed, make it more difficult for \camera{some} older adults to learn to use computers~\cite{Echt1998EFFECTS}. \textit{Physical declines} were also suggested as barriers for some older adults to use tablets including poor eyesight, arthritis, and fractured wrist and fingers~\cite{vaportzis2017older}. 
Moreover, Rama et al. pointed out \rv{some} older adults \rv{such as those experiencing age-related cognitive declines} face challenges when learning \textit{the new generation of UIs} since they may not effectively transfer the knowledge of older technologies to the current UIs which are quite different~\cite{Rama2001Tech}. Furthermore, \camera{some} older adults expressed concerns about the lack of effective instructions and guidance when using tablets and technology, and didn't have sufficient knowledge and confidence~\cite{vaportzis2017older}.

Specifically, on smartphone usage, \camera{some} older adults face similar challenges with other technologies such as computers and tablets. \textit{Aging cognitive ability} such as memory decline makes it difficult to remember the instructions over time, and \textit{aging physical ability} like mild vision and mild motor impairment such as difficulty in finger movement makes \camera{them} hard to see clearly and operate on the screen~\cite{Mohadis2014ASUB, Pang2015OAADS}. Moreover, \textit{unfamiliarity with smartphone usage and the related terminology, the lack of knowledge about the smartphone, and the accompanying low self-efficacy} make \camera{them} reluctant and unsure to operate smartphones~\cite{HowUnfamiliar2013Ishihara, Mohadis2014ASUB, Pang2015OAADS}. It also makes \camera{some} older adults feel confused once they encountered unclear instructions and unintentional gestures~\cite{Kobayashi2011Elderly}. Furthermore, the \textit{lack of social support and resources} to learn hinder the appropriation of smartphones among older adults~\cite{Pang2015OAADS}. In the meantime, those challenges would lower \camera{their} intention to learn and significantly influence their adoption on smartphones~\cite{Kim2016Acceptance}.

\subsubsection{Learning Needs and Learning Preferences.}
Leung et al. investigated that older adults value the exact steps to complete a task more than knowing how the software works~\cite{Leung2012}. A more recent semi-structured interview conducted by Pang et al. echoed this finding that older adults prefer \textit{step-by-step} instructions~\cite{Pang2021Association}. Previous work has suggested that older adults prefer using an instruction manual (e.g., textual guidance) to conduct trial-and-error, though complaining it is difficult to use due to the issues such as lack of detail and hard mapping~\cite{Leung2012, HowUnfamiliar2013Ishihara,fan2018guidelines}. Device help features (e.g., text description, interactive tutorial, video inside the applications), especially interactivity ~\cite{Pang2021Association}, were also found to be preferred by older adults due to the given right direction to follow and convenience to access but were criticized for containing content that is hard to understand~\cite{Leung2012}. As new technologies develop and the demographics of aging shift, older adults were found to be less reliant recently on the instructional materials than in the past due to the complexity of the language, instead, they \textit{prefer a more independent and flexible learning approach} as well as \textit{remote support}~\cite{Pang2021Association}. Trial-and-error is a typical way to learn independently in many domains~\cite{kavanaugh2019just}, and has also been proven to have benefits for older adults in source memory and episodic memory~\cite{cyr2012trial, cyr2015mistakes}. Although \camera{some} older adults face challenges when doing trial-and-error such as lack of encouragement for self-diagnosis~\cite{fan2018guidelines}, and showing less preference in using trial-and-error~\cite{mitzner2008older, Leung2012}, previous research suggested that older adults may be more willing to try if they are supported to be more confident that errors would not lead to severe consequences~\cite{barnard2013learning}. 

Researchers revealed that \camera{some} older adults feel concerned about troubling others and prefer learning alone at their own pace or with their peers to avoid embarrassment~\cite{Si2009Desiring, Leung2012}. Recent research found that older adults enjoy the experience of learning collaboratively with their spouses, another family
members, and those within their own social network ~\cite{Pang2021Association}. This difference could be that the collaborative approach in the latter study is more related to remote help \rv{via video chat} and has more efficient tools for guiding \rv{with a variety of supports including an integrated Google Search, feedback, step-by-step video instructions}, which reduces the embarrassment and the concern of troubling others\rv{~\cite{Lindley2009Desiring}}, and further implies the benefit of remote help.




\subsection{Remote Assistance on Using a Smartphone}

Live screen sharing applications \rv{(e.g., InkWire~\cite{InkWire}, TeamViewer~\cite{TeamViewer}, or AirDroid~\cite{AirDroid})} could provide real-time assistance by controlling the Android device but highly depends on the stable and high bandwidth network connection and may violate the user’s privacy as well. Moreover, it adds an extra burden if the help giver and the help seeker have different schedules or if the help seeker has a poor memory and asks for support repetitively, which may happen commonly in the older help seekers. Vyas et al. further developed the remote assistance with an app called APPAmigo by capturing and recording the events from one device and recreating them on the other device to avoid requiring complete access to control the entire remote device~\cite{Vyas2019AppAmigo}. While informative, the approach focus on directly solving the problem instead of motivating the user to learn by themselves. A potential risk of it is that the user may depend on help-givers gradually and still lack of ability to perform similar tasks by themselves. Therefore, a more appropriate approach is providing support to help older adults learn independently. 

Tutorials are useful learning materials to provide step-by-step guidance on how to perform a task.
Exploiting text-based tutorials and video tutorials are the common ways when people learn how to use certain applications on smartphones. However, static instructions may not include sufficient detailed information. While video tutorials contain all the necessary detail, they are hard to follow for the novice due to the quick speed and the tedium of constantly switching between tutorials and apps. Pause and Play links the important events with the video to automatically pause and play the video to stay in sync with the user~\cite{Pongnumkul2011}. MixT combined the static and video clips together to provide more effective tutorials~\cite{Chi2012Mix}. While users still need to match the content of the tutorial with the real application by themselves. Contextual tools like Sketch-sketch Revolution offer in-application interactive tutorials~\cite{Jennifer2011Sketch-Sketch}, DemoWiz provides within-context features, which have proven to be effective for novice users~\cite{Chi2014DemoWizRS}. However, most of them are application-dependent. Furthermore, Wang et al. developed EverTutor which could provide a contextual interactive step-by-step guidance on smartphones by demonstration universally~\cite{Wang2014EverTutor}. Similarly, a more recent design Remo proposed by Aftab et al. generates interactive tutorials to provide remote help asynchronously as a web browser extension by demonstration~\cite{Remo}. However, they are limited in visual-based guidance and lack consideration of older adults' needs. Moreover, they are designed as a guiding tool instead of the one supporting and motivating older adults to learn smartphones. Synapse, as an App agonistic service, provides two modes of interactive support. Synapse (Basic mode) provides intuitive interactive support by highlighting all the components to click on, and Synapse (Trial-and-Error mode) provides instant support and error recovery support with voice assistance to promote older adults learning.  

\section{Key Design Features}

\begin{figure*}[htb!]
  \centering
  \includegraphics[width=\linewidth]{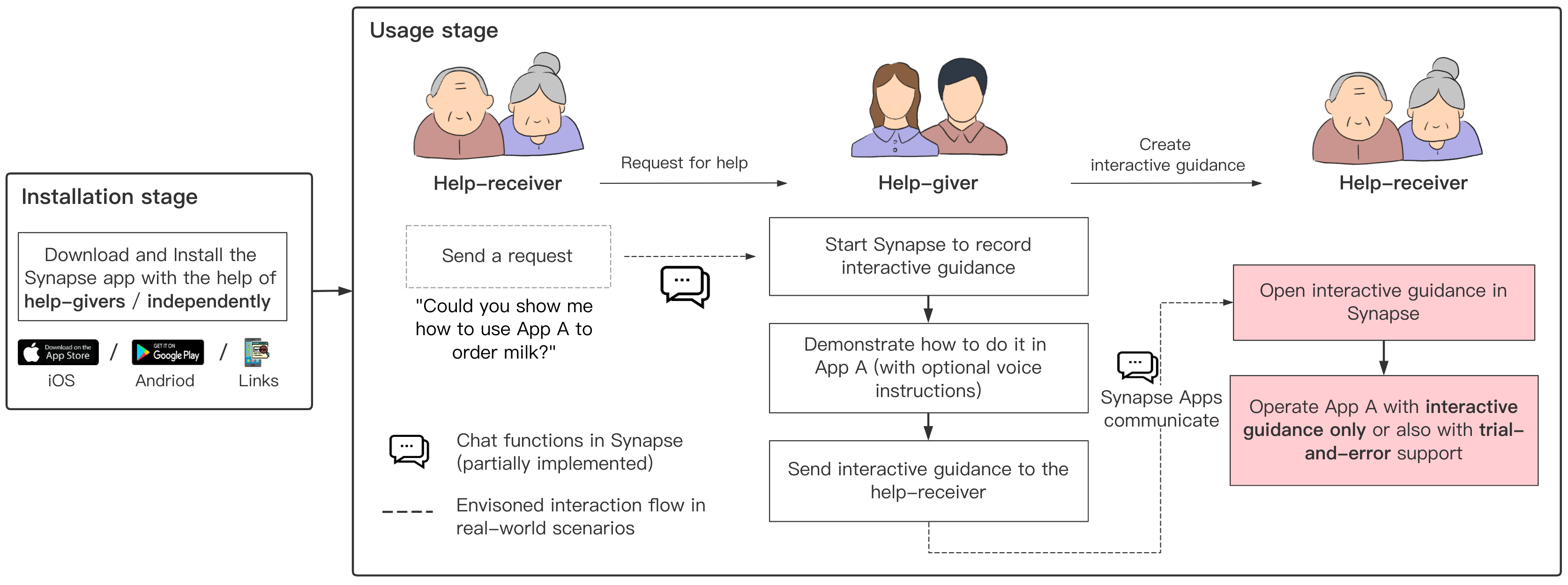}
  \caption{The whole interaction flows between help-givers and help-receivers when using Synapse. Older adults first download and install Synapse either by themselves or with the help of help-givers. In the Usage stage, a help-receiver sends a request (e.g., How to use App A to order milk?) to \camera{their} help-giver via the chat functions within Synapse or other channels (e.g., phone calls). The help-giver starts Synapse on his phone and starts to demonstrate how to use App A to complete the task step-by-step. \camera{They} could also add voice instructions to clarify steps. Synapse records \camera{their} interactions in App A and the voice instructions as a tutorial script. Then, \camera{they} sends the help-receiver the script, who will be run by the Synapse on \camera{their} phone and provides interactive guidance (Basic mode) and trial-and-error support (Trial-and-Error mode) to guide \camera{them} to use App A. This work focused on understanding older adults' performance and user experience of using interactive guidance and trial-and-support compared to traditional video tutorial \camera{(evaluating the highlighted procedure.)}}
  \Description{Older adults first download and install the Synapse app with the help of help-givers or independently. When older adults encounter an issue (e.g., how to use App A to order milk), they can send a request to help-givers either via the Synapse app or via other methods like phone call. After receiving the request, help-givers start Synapse on their phones to record interactive guidance. They can simply record the usage process of APP A with optional voice instructions by demonstration and send it to older adults. Older adults can see the guidance file transferred through the chat system, and select it, then choose using interactive guidance only or with trial-and-error support}
  \label{fig:systemUsage}
\end{figure*}

\subsection{Interaction Flow}
\label{sec:interactionflow}
\rv{Fig.~\ref{fig:systemUsage} shows the whole interaction flows for both help-givers and help-receivers when using Synapse service. In the Installation stage, older adults first download and install Synapse from app stores or shared links either by themselves or with the help of help-givers. In the Usage stage, when a help-receiver does not know how to use App A to complete a task (e.g., order milk), she sends a request to a help-giver (e.g., children, friends) via the chat functions within Synapse or other channels, such as phone calls. The help-giver starts Synapse on his phone, which runs in the background, and starts to demonstrate how to use App A to complete the task step-by-step. He could also add voice instructions to clarify the steps. Synapse records his interactions and his voice instructions as a tutorial script. Then, he sends the tutorial script to the older adult via Synapse. When the older adult clicks the tutorial and selects the mode (Basic or Trial-and-Error) to use, Synapse runs it at the background and provides the recorded interactive guidance and trial-and-error support depending on the selected mode to guide her in using App A.}

\rv{Our research question (RQ) was to understand older adults' performance and user experience of using interactive guidance and trial-and-error support compared to the traditional video tutorial. To better answer RQ, we focused on the implementation and evaluation of interactive guidance and trial-and-error support for older adults (i.e., help-receivers) with a minimal implementation on the help-giver side and the communication between the two sides, which are indicated with dash lines in Figure~\ref{fig:systemUsage}. In Sections~\ref{sec:realworld} and ~\ref{sec:limt}, we will further discuss how to make Synapse work in real world scenarios and important future research directions. Next, we present the key design features for the help-giver and the help-receiver with an emphasis on the latter in this work.}


\subsection{Help-giver}
\label{sec:help-giver}

\begin{figure*}[htb!]
  \centering
  \includegraphics[width=8cm]{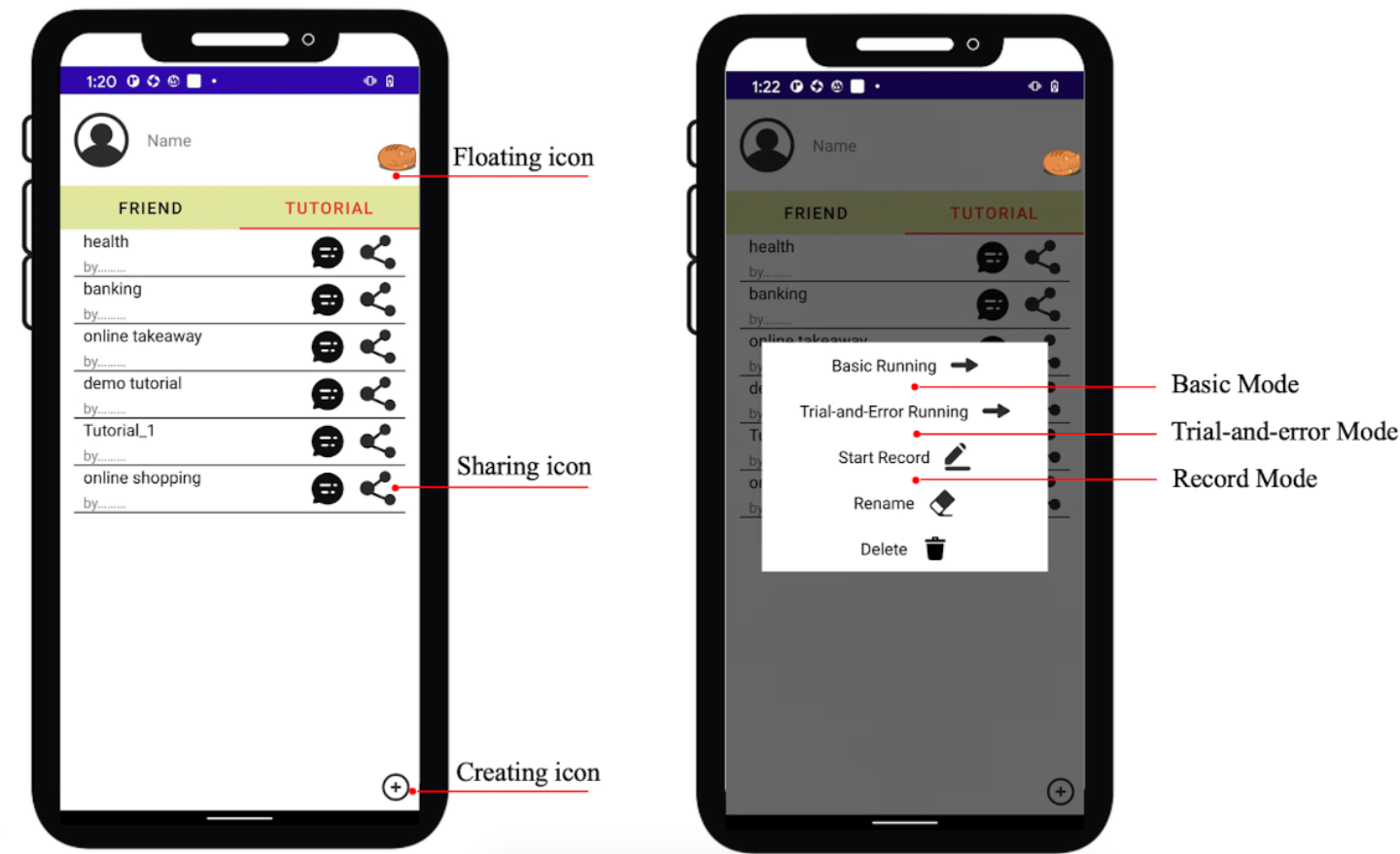}
  \caption{Overview of Synapse. The left one shows the "TUTORIAL" tab view UI of Synapse by default. The interface contains two tab view. "FRIEND" means the chat function, "TUTORIAL" lists all the scripts with sharing icon next to them. The right one shows the pop-up window when selecting a certain file, including the options of Basic Mode, Trial-and-Error Mode and Record Mode.}
  \Description{}
  \label{fig:overview_process}
\end{figure*}

\begin{figure*}[htb!]
  \centering
  \includegraphics[width=\linewidth]{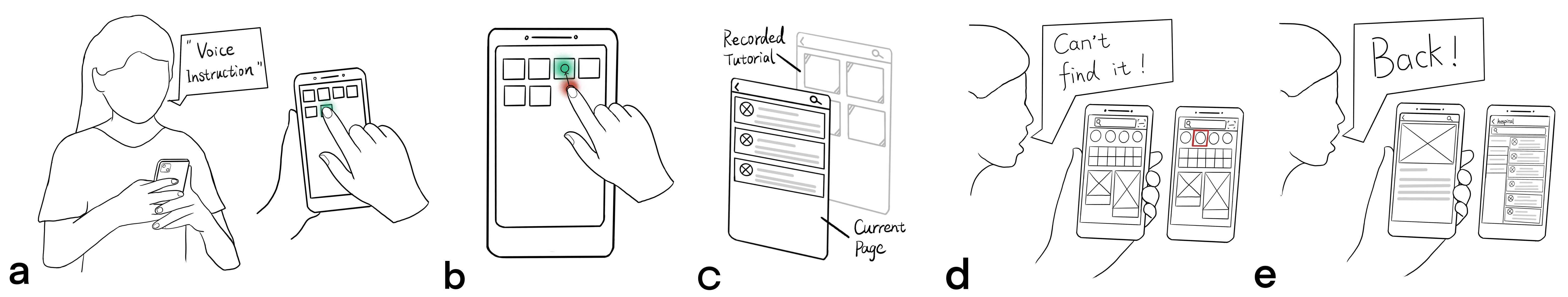}
  \caption{Design features. (a) shows a help-giver creating tutorial. (b) shows when the user clicks outside the view, Synapse would help to automatically calibrate to the nearest view. (c) shows that a robust enhancement mechanism to check if the current click position matches with the one on the recorded page. (d) shows the trial-and-error support with voice assistance. (e) shows the error recovery support with voice assistance.}
  \Description{}
  \label{fig:Dfeature}
\end{figure*}

\textbf{Create tutorial}.
Previous research generated context tutorials by writing scripts that add more constraints and burdens to the tutorial creators~\cite{yeh2009sikuli, yeh2011creating}. Drawing inspiration from the work such as EverTutor~\cite{Wang2014EverTutor} and Remo~\cite{Remo}, we allow \rv{help-givers} to create their own tutorials by simply demonstrating and recording the bounding box coordinates, package name, class name, and description text of the icon on the screen that the user clicked on, and save it as a node for matching when using the tutorial. \rv{Fig.~\ref{fig:overview_process} shows the overview of Synapse. Help-givers} can create a tutorial by clicking the Add button \rv{at the bottom right corner of Fig.~\ref{fig:overview_process}}, entering the tutorial name, and clicking OK. Then, we allow them to start recording a tutorial at any time with voice instructions, providing them with a higher degree of freedom as Fig.~\ref{fig:Dfeature} a shows.

When the user is ready to start recording, they can click the floating icon on the screen as Fig.~\ref{fig:overview_process} shows and select "Start recording tutorial". \rv{The user's gestures are detected through Accessibility Service "GESTURE\_TOUCH\_EXPLORATION" including single tap, double tap, swipe, etc~\cite{AccessibilityAPI}, and typing gestures can be fetched through Accessibility.NodeInfo.isTextEntryKey. While our implementation can support all these gestures, we included only single tap, which is the most common operation people use in daily life, in our user study as our focus was not testing how older adults use different gestures but on evaluating their performance and user experience of using interactive guidance and trial-and-error support. Nonetheless, we acknowledge and discuss this limitation in Section~\ref{sec:limt}.} When this recording is successful, the user will be provided with feedback on the screen, informing the user that the recording is successful. In addition, to make the tutorial easier to be understood by users, we allow users to \textit{record voice instructions} for each step of the tutorial. The user can record the voice by clicking "Start Voice Recording" in the menu before each click, the program will call MediaRecorder to record voice instructions and store them as a .amr file. When the user finishes recording the tutorial, click "End Recording" and the program will save all the recorded node information as a JSON text file. Then the user can view the recorded tutorials in the tutorial list and could share them with friends by clicking on the "share button".

\subsection{Help-receiver}
\label{sec:helpreceiver}
\textbf{Interactive guidance}.
We create an overlay over the screen and highlight the location on the screen that needs to be clicked by parsing the bounding box (bbox) coordinates from the JSON file, as shown in Fig.~\ref{fig:teaser} a1, a2. In the meantime, the voice instruction recorded by the help giver will play to give an explanation of the current step. When the user clicks, the program will record the click's coordinates on the screen. Then the program needs to judge the location relationship between the click coordinates and the bbox coordinates. When the click is detected outside the bbox, the program will not perform any action to protect the user from a potential wrong click. When the click is detected inside the bbox, we use the dispatchGesture function~\cite{Accessibility} provided by the Accessibility Service~\cite{Accessibility, Rodrigures2018Aidme} to simulate the user's click on the corresponding location and trigger the corresponding action. We also enable users to terminate the running tutorial at any time in case they would like to try the following steps by themselves. When the entire tutorial is completed, the program will exit the running state and the interface will return to any clickable mode, and the user will be prompted that the tutorial is completed. 

\textit{Precise click calibration}.
As applications become more complicated these days, there are often multiple UI components crammed into one interface making it crowded including small icons which are hard to click on precisely especially for older adults~\cite{Jin2007Touch}. Therefore to improve the correct rate of clicks, we design an algorithm to discriminate which icon is the one that the user is meant to click on as Fig.~\ref{fig:Dfeature} b shows. We first model the current page as a tree structure and get the root node with AccessibilityWindowInfo. Then we recursively search all AccesssibilityNodes (refers to each view) on the page starting from the root node and extract the bbox coordinates for the following calculation. We calculate all the distances between the center coordinates of the view and the coordinates of the position the user clicks on and find the closet view. Finally, we passed the center coordinates of the view to the dispatchGesture function and execute the simulated click. \rv{It is worth mentioning that this function would only be activated in the Basic Mode, which aims to help older adults finish specific tasks smoothly. It is not activated in Trial-and-Error Mode, which aims to encourage older adults to learn independently.}

\textit{Robustness enhancement}.
We design a robust enhancement mechanism to avoid unexpected mistakes happening during usage. For example, when Accessibility Service may not fully record the users' click operation accidentally, there is a potential risk that while the app turns to the next page, the tutorial still stays in the previous step. That may lead to the inconsistency between the tutorial and the app page. Therefore, we detect whether the position in the recorded tutorial matches on the current page or not for each operation which could cause a page jump as Fig.~\ref{fig:Dfeature} c shows. If not, then the system would prompt the error to let the user manually fix the inconsistency. \rv{For example, the user could come back to the last page manually to keep the page consistent with the one in the tutorial's step.} After the user fixes the problem, they can click on the floating icon and choose "resume" to continue running the tutorial. 

\textbf{Trial-and-Error}.
To provide more flexibility in learning in Trial-and-Error mode, we allow the user to click on any position. The interface will changes normally, but if the user clicks on the corresponding position in the tutorial, the user will be prompted with the correct click as Fig.~\ref{fig:teaser} b5 shows, otherwise, the user will be prompted with the wrong click as Fig.~\ref{fig:teaser} b2 shows. We only prompt the user when they deviate from the correct position for the first time, and will not prompt the user if they continue to click on the wrong position based on the feedback from the pilot study from 4 older adults. They felt annoyed if they were continuously notified that they were in the wrong flow since they totally knew this with the first prompt and just would like to continue trying.

Considering the difficulties that may exist for older adults when operating their phones, we allow the user to trigger the corresponding action by natural conversation. The speech recognition module is constantly on, and the user can input speech at any moment without any time limit. When users can't find where to click, they can trigger the highlight rectangle by saying sentences containing "can't find it" as Fig.~\ref{fig:Dfeature} d shows. When the keyword "can't find it" is parsed, the program will start the same overlay as the one in the basic mode, the user can only click on the highlight position on the overlay to go to the next step, when the user successfully goes to the next step, the overlay disappears and the user can still click freely.

\textbf{Error recovery support}.
To better support the user conducting trial-and-error, we design an error recovery mechanism applying voice assistance to support the user going back and starting over the tutorial. When the user wants to go back to the previous level menu because of a wrong click or other reasons,  they can go back to the previous level page by saying "back" as Fig.~\ref{fig:Dfeature} e shows. When the user thinks they have made too many mistakes or want to restart the tutorial, they can say "start over" to return to the main interface of the phone and reset the progress of the tutorial. 

\textit{Execution}.
When the "back" keyword is resolved, the program checks the status of the top element of the stack and if the last click is "correct", the current running tutorial node will be back once; otherwise, it will remain. After triggering "back", based on the Accessibility Service, the system will return to the previous level page. When the user says "start over", it will jump directly to the main interface and reset the currently running tutorial to the initial node.

\textit{Back to the right page}. 
We set up a stack to record each time the user clicks correctly or not. When the user clicks correctly, we get the Id of the root window where the clicked UI element exists from the Android Accessibility Service. Then we check if the top element of the stack is the same as the current window Id. If they are different, we push 1 into the stack and empty all the 0s in front of it; otherwise, we push 0. When the "back" voice is detected, the program checks and pops the value of the top element of the stack, and if it is 1, then the node index of the tutorial is subtracted by 1, which means that the progress of the tutorial is set back one frame and the user is prompted to return to the correct page. If it is 0, the tutorial progress is kept unchanged. In this way, when the user is back to the right page, the user will get a prompt that says that the current page is the right page. Therefore, the user can be notified when and where to stop on the right page.


\section{Implementation}

\textbf{Hardware Setup}.
We have built our app on Google Pixel 4, and Android 11, and tested it to work on Android phones up to 7.0. The app does not require root access and is based on Java 1.15.0. The tutorials and soundtracks of the user's personal record are stored on the phone, and we use a dedicated server to store the tutorials uploaded by different users in a zip file.



\textbf{Background Accessibility Service}.
We use Android's Accessibility Service to support listening and control operations on the backend, the purpose of the Accessibility Service is to help users with visual, physical, or age-related limitations to use Android devices and applications more easily. After the service is started, when the text or position of a view changes, or when the user triggers an operation such as click, long press, swipe, etc., the service will receive the Accessibility Event and trigger the corresponding action.

\textbf{Service Action with Different State}.
We set 4 states for the system: Normal, Recording, Running, and Pause. When the user starts recording, the system state enters into the Recording state, at this time, Accessibility Service keeps listening to the user's click on the view and records the corresponding view's bbox, package name, class name, and text description of the corresponding view. When the user starts running the tutorial, the system state enters into the Running state. If the user is using basic mode, Accessibility Service will not process any Accessibility event. If the user is in Trial-and-Error mode, Accessibility Service will process the "clicked" Accessibility event and compare it with the node information stored in the running tutorial to determine whether the user clicked correctly and give the corresponding feedback.

\textbf{Tutorial File Encode/Decode}.
We used Google's Gson package to store and read the tutorial.
For each view clicked by the user, we record the bbox coordinates, package name, class name, text description, and soundtrack file path of the corresponding view, if there is a recording. We construct the TutorialEntity data type to package this information into a node so that the entire tutorial consists of multiple nodes. When the recording is complete, we create a JSON file with the same name, based on the tutorial name, and use the Gson package to write the list of nodes to the JSON file.
When the user enters the running state, the system will automatically load the JSON file corresponding to the tutorial, use the Gson package to reduce the node information in the file to a List of nodes, and create the tutorial data type based on the nodes to support the entire running state.

\textbf{ASR (Automatic Speech Recognition)}.
We use iFLYTEK\'s speech dictation SDK\footnote{\url{https://www.xfyun.cn/services/online_tts}} to support our keyword speech recognition feature. The SDK can support Mandarin and common Chinese local dialect recognition which is helpful since we conducted the user study in China. At present, it is designed to support recognizing the sentences containing "can\'t find it", "back", and "start over" keywords. In Trial-and-Error mode, the ASR module stays on all the time and pauses listening only when the user triggers "can\'t find it". When the ASR module listens to the speech input, it returns the recognition result in its onResult method through the parsing of its algorithm model.



\section{User Study}
We conducted a user study with older adults who are 60 and over to compare the guidance and learning effectiveness between the traditional screen recording approach, Basic mode, and Trial-and-Error mode of "Synapse". \rv{As we have explained in Section~\ref{sec:interactionflow}, this work focused on understanding the performance and user experience of using interactive guidance and trial-and-error support by older adults as help-receivers, we did not involve help-givers in the user study. Instead, in the user study, we attached the same voice instructions to the Synapse tutorial (both Basic and Trial-and-Error modes) and the Video tutorial to simulate that help-givers have already generated the corresponding helping materials. We simulated that older adults have already received the tutorials and started our user study by using the tutorials. we will discuss important future work involving help-givers in Sections~\ref{sec:realworld} and ~\ref{sec:limt}.}

\subsection{Participants}
We recruited 18 older adult participants through local community centers, nursing homes, and social media platforms. \camera{The inclusion criteria included 1) being aged 60 or over; 2) having the basic knowledge of using smartphones without being an expert; 3) being in a healthy condition that allows them to participate in the study independently.} 
Table~\ref{tab:demographic} shows their demographic information. In sum, they
were aged between 60 and 74 ($Median = 64, SD = 4$). Twelve were female and six were male. \camera{We administered a mobile literacy questionnaire to evaluate their basic mobile phone usage. All could pass at least 10 of the 14 items in the questionnaire}. \camera{While some participants reported to have age-related memory decline (P5, P13) or vision decline (P7, P9), all were in a generally healthy condition and were able to participate in our study independently.} The user study sessions lasted for 60-90 minutes, and participants received \$15 for their participation.

\begin{table}[htb!]
    \caption{Participants' demographic information}
    \label{tab:demographic}
    \Description{Table \label{tab:dempgraphic} demonstrates the age, sex, location, and prior device usage experiences, such as computer, android phone, iPad, and iPhone of 16 participants each.}
    \begin{tabular}{c|c|c|c|c|c}
    \hline
    \rowcolor[gray]{0.9} Id & Age & Sex & \rv{Recruit resource}  & Years of smartphone usage & \makecell[c]{Self-estimated number of hours \\using smartphones per day} \\
    P1 & 63 & F & \rv{nursing homes} & 10 & [2, 4) \\
    \hline  
    P2 & 66 & F & \rv{social media platforms} & 5.5 & [8, 12) \\
    \hline 
    P3 & 63 & M & \rv{social media platforms} & 4 & [4, 8) \\
    \hline 
    P4 & 61 & F & \rv{social media platforms} & 6 & [4, 8) \\
    \hline  
    P5 & 64 & F & \rv{social media platforms} & 10 & [8, 12) \\
    \hline 
    P6 & 64 & M & \rv{social media platforms} & 10 & [8, 12) \\
    \hline 
    P7 & 74 & F & \rv{social media platforms} & 4 & [2, 4) \\
    \hline 
    P8 & 68 & F & \rv{social media platforms} & 4 & [8, 12) \\
    \hline 
    P9 & 69 & F & \rv{social media platforms} & 5 & [8, 12) \\
    \hline  
    P10 & 60 & M & \rv{social media platforms} & 5 & [2, 4) \\
    \hline 
    P11 & 63 & M & \rv{local community centers} & 10 & [2, 4) \\
    \hline 
    P12 & 65 & F & \rv{local community centers} & 10 & [2, 4)  \\
    \hline 
    P13 & 61 & F & \rv{local community centers} & 10 & [4, 8) \\
    \hline
    P14 & 67 & M & \rv{local community centers} & 10 & [4, 8) \\
    \hline
    P15 & 65 & F & \rv{local community centers} & 15 & [4, 8) \\
    \hline
    P16 & 61 & M & \rv{social media platforms} & 8 & [8, 12) \\
    \hline
    P17 & 64 & F & \rv{local community centers} & 8 & [4, 8) \\
    \hline
    P18 & 60 & F & \rv{social media platforms} & 5 & [4, 8) \\
    \hline
    \end{tabular}
\end{table}

\subsection{Design}
We considered Condition as a within-subject factor. To mitigate the learning effect between different conditions, we prerecorded tutorials with three different apps corresponding to three conditions, \rv{which are the Basic mode of Synapse, the Trial-and-Error mode of Synapse, and the video approach. The video tutorial was a screen recording with narration to simulate the natural way of help givers typically do in their daily lives and the length is 1 min.} These three apps are about health, finance, and online food delivery, which could be used in older adults' daily life. In the choice of apps, we chose less popular ones and made sure participants had never used them before the experiment. \rv{We fixed the apps' order and assigned} each participant to one of the six combination condition status orders, following a balanced Latin square \rv{to mitigate the effect brought by the app and the experimental order of conditions}. 

\subsection{Procedure}
The experiment was designed to fit into two sessions. In session 1, participants were asked to complete three tasks, each with different support (screen recording video, Synapse (Basic), Synapse (Trial-and-Error) separately on three apps to avoid the learning effect from the previous condition. In session 2, participants were asked to complete three similar tasks without support on these three apps. For example, making a hospital appointment using the telemedicine app in dentistry instead of making one in ophthalmology. The motivation for including session 2 is that we would like to identify whether the condition of participants learned in session 1 would have any significant influence on their performance in session 2. As the start, participants were asked to complete a brief questionnaire about their demographic information. Participants then completed the experiment tasks in session 1. They were introduced to using "Synapse" with sample tutorials and shown that (1) they can click the position where the red rectangle refers and are not allowed to click other positions in the basic code, and (2) they could say sentences containing "could not find" to trigger the app to highlight the position to click on, say "back" to return to the previous step and "start over" to start from the beginning point in the trial and error mode. We also let them try using the sample tutorials by themselves to get familiar with the use of "Synapse". (They were only introduced to a certain mode before the certain condition usage to avoid the cognitive effect.) \rv{When trying with video approach, participants were not constrained with the strategies on how they leveraged the video tutorial. They could either play the video in parallel while conducting the task or look through to get a general overview of it.} Following each condition, participants completed a 7-point Likert scale questionnaire (a modified System Usability Scale questionnaire) about that condition and were asked several questions including whether the current condition actually helps, if yes, what features work, what challenges still bother them. After that, at the end of session 1, participants were asked about the comparison of these three conditions, and further improvements. Finally, participants were asked to complete three learning tasks one by one in session 2 on the apps that they learned in session 1 to evaluate their learning performance.

\subsection{Measures and Data Analysis}

For task performance, we measured the time for completing a task and the number of mistakes they made. Mistakes here mean the operation of clicking on the wrong icon which would either lead the participant to the wrong page or stuck on the current page. We additionally included the result of the 7-point Likert scale user experience questionnaire as a measure to evaluate the user experience. 

For objective quantitative data, we computed descriptive statistics and performed mixed-effects regression models, which will be described in detail in Section 6.1. For subjective qualitative data, we performed the Friedman's test on the subjective result of the 7-point Likert scale questionnaire and for whose show a significant result, we performed the post-hoc test to specify which pair cause the result. For qualitative data, two researchers of the team looked through the video and wrote down observation points such as where participants made mistakes and what kind of mistakes made them hard to recover as well as coded the interview answers. Then they performed thematic analysis independently and discussed the common themes that emerged from the texts. Finally, the themes were further discussed with one additional researcher on the team and consolidated into the key findings, which will be described in detail in Section 6.2.   

\section{Results}

We first present quantitative results from the experiment and subjective 7-point Likert scale questionnaire and then present qualitative results including interviews and observations. 

\subsection{Quantitative Results}



\subsubsection{Task Completion Time} 
\begin{figure*}[htb!]
  \centering
  \includegraphics[width=12cm]{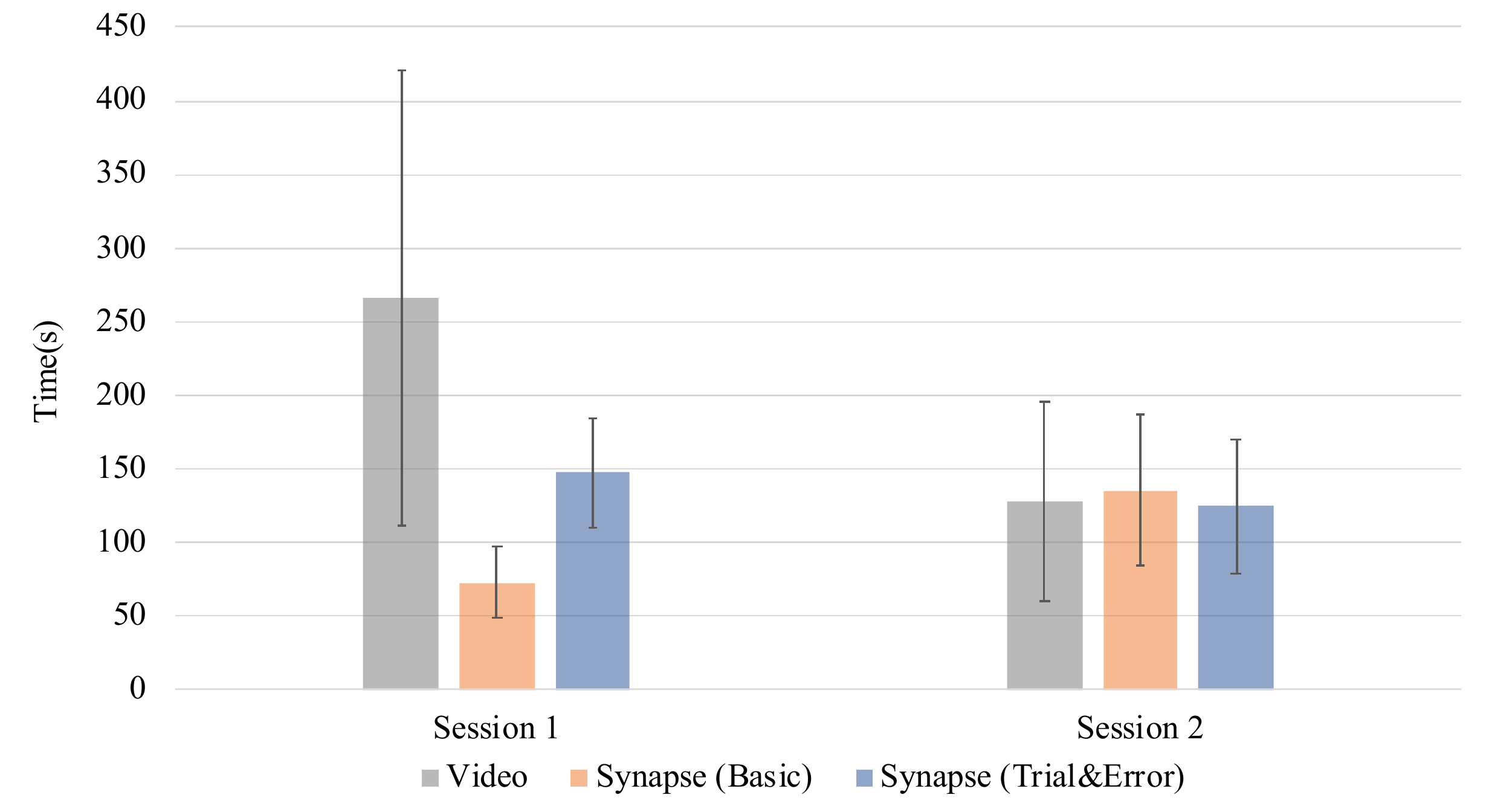}
  \caption{The mean and standard deviation of time spent in session 1 and session 2 on each condition}
  \Description{}
  \label{fig:time}
\end{figure*}

We calculated the mean and standard deviation of the dependent variable time spent on each condition during session 1 to indicate the efficiency with each support and session 2 to indicate the corresponding learning effect. 
Fig.~\ref{fig:time} shows the mean and standard deviation of time spent on each condition during session 1 and session 2. Participants spent the significantly least time with the interactive (Basic) support during session 1 which is 73.3 s (24.5), while spent the least time with the interactive (Trial-and-Error) support during session 2, which is 124.8 s (45.4), though not significantly. The result shows that the interactive (Basic) support saves participants' time significantly, and the interactive (Trial-and-Error) support slightly improves \camera{our participants}' learning performance. 

For time spent in session 1 and in session 2, we further performed a separate mixed-effects regression with \textit{condition status} as the fixed effect and \textit{participants} as the random effects. The regression model further included the participants’ self-reported mobile experience(e.g., years using smartphones, hours spent on smartphones per day, things using smartphones to do), education level, age, and gender as control variables. For each test, we removed outliers with the dependant variable as outside 1.5 times the interquartile range and made sure there was no multicollinearity (all VIF < 3). We found that Synapse (Basic) ($\beta$ = -100.2, SE = 15.7, F(2, 27) = 26.1, p < .001) have significance contrast with the video on the dependent variable time spent in session 1, while Synapse (Trial-and-Error) ($\beta$ = -26.0, SE = 15.7, F(2, 27) = 26.1, p = .1) shows the trending toward significance. Interestingly for session 2, none of the factors have a significant effect on the dependent variable time spent in session 2. 
\subsubsection{Number of Mistakes}
\rv{We counted} the total number of mistakes participants made during the session 2 \rv{in each condition, and the result} shows that participants made the fewest mistakes (n = 10) on the app which they learned in session 1 using Synapse (Trial-and-Error), followed by the video approach (n = 13) and Synapse (Basic) (n = 23). \rv{One} possible reason \rv{why participants made the most mistakes with Synapse (Basic)} could be that participants did not think so much with direct guidance in Synapse (Basic) during Session 1. For dependent variable the number of mistakes made by participants in session 2, we found significant contrast between video and Synapse (Basic) ($\beta$ = 0.7, SE = 0.2, F(2, 31) = 5.9, p < .01), while no significance between video and Synapse (Trial-and-Error). For all of these three dependent variables, none of the demographic factors were found to have a significant effect on it.

\subsubsection{7-point Likert Scale User Experience Questionnaire.} 
\begin{figure*}[htb!]
  \centering
  \includegraphics[width=\linewidth]{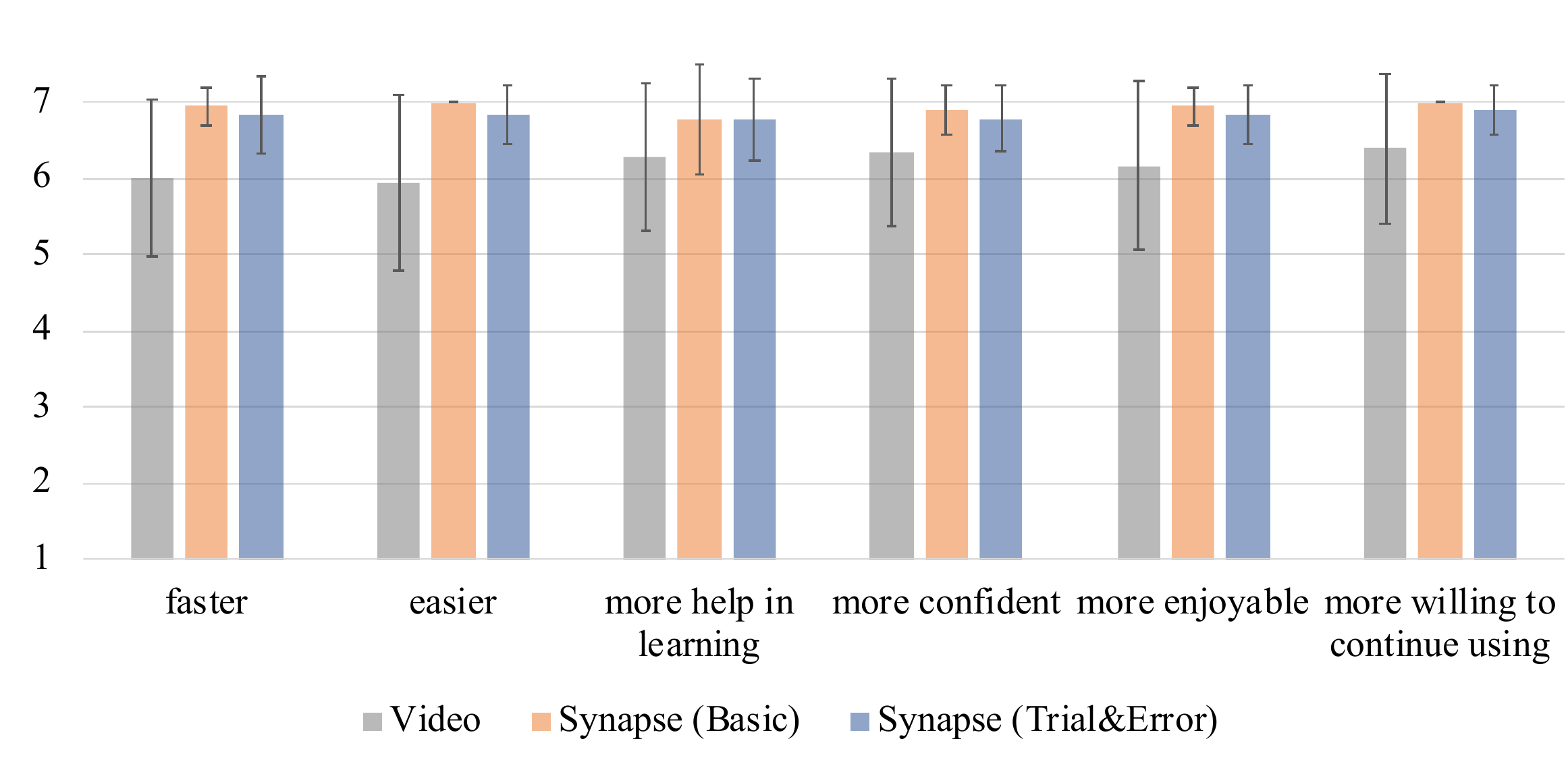}
  \caption{The mean score of the 7-point Likert scale user experience questionnaire on each condition}
  \Description{}
  \label{fig:SUS}
\end{figure*}

We first calculated the mean and standard deviation of scores obtained from the 7-point Likert scale questionnaire. Fig.~\ref{fig:SUS} shows the mean score and standard deviation of the result. Generally, participants rate higher scores for both two modes of the interactive approach than for the traditional video one, while rated similar between the two modes of the interactive approach. It indicates that participants feel faster, easier, more help in learning, more confident, more enjoyable, and more willing to continue using the interactive approach Synapse than the traditional video one. 

Following that, we performed Friedman's test on all these 6 levels of user experience. We found that feeling faster (p < .001), easier (p < .001), more confident (p < .05), more enjoyable (p< .001), and  willing to continue to use (p < .05) have significant relationship with the condition status. We then performed the post-hoc test on these 6 levels separately, and found out that both the two interactive modes of "Synapse" have a significant relationship with the traditional video approach on the feeling of completing the task faster (p1<.001, p2 <.01), easier (p1, p2 < .001), enjoyable (p1 <.01, p2 <.05). For feeling more help in learning, the two show trending towards significance (p1 = .06, p2 = .09) with the traditional video approach. For feeling more confident, interactive (basic) also shows trending towards significance (p = .09) with the video one. For willingness to continue use, the interactive approach (Basic) is significantly better than the video (p < .05), while the interactive (Trial-and-Error) one shows the trending towards significance (p = .08). One of the possible reasons for showing the trending toward significance rather than the real significance between the interactive ones and the video one could be the unfamiliarity of the prototype since this kind of interactive approach is a new thing while the video is not.

\subsection{Qualitative Results}
We listed the key pros and cons of the three approaches as Table.~\ref{tab:proscons} shows. Among these three approaches, there are 11 participants who prefer to use the Synapse (Trial-and-Error mode), 6 prefer to use the Synapse (Basic mode) and 1 had no preference between the two. First of all, participants stated that they would like to be independent to learn to use smartphones. They felt that all of these approaches would help them to some extent when their children are not available or getting impatient with their slow speed of learning. 

\begin{table}[htb!]
    \caption{Pros and Cons of videos, Synapse (Basic), Synapse (Trial-and-Error)}
    \label{tab:proscons}
    \Description{Table \label{tab:proscons} demonstrates the pros and cons of these three approaches.}
    \begin{tabular}{c|c|c}
    \hline
    \rowcolor[gray]{0.9} & Pros & Cons \\
    \hline 
    \cellcolor[gray]{0.95}Video & \makecell[c]{voice instructions, \\awareness of the entire process} & \makecell[c]{lack of the control over speed, \\lack of the highlighting components}  \\
    \hline  
    \cellcolor[gray]{0.95}Synapse (Basic) & \makecell[c]{voice instruction, time saving, \\low leaning barrier, correctness} & \makecell[c]{\rv{restrictions on clickable areas}, \\ limited performance of visualization}  \\
    \hline 
    \cellcolor[gray]{0.95}Synapse (Trial-and-Error) & \makecell[c]{conformation to natural usage habits,\\ visual and audio feedback, \\ promotion to learn, error recovery support} & \makecell[c]{limited performance of visualization, \\limitated time saving} \\
    \hline 
    \end{tabular}
\end{table}
\subsubsection{\rv{Features} of Video Approach}
The main preferable feature of the video approach is the \textbf{voice instructions} from the video. 
``\textit{It contains the voice instruction which is great. I could know more about the current operation.}''- P1. P6 even commented that voice is better than video since he did not know where to click on it by watching the video anyway. 
\rv{We observed an interesting phenomenon that participants chose different strategies for watching video tutorials. Some of them chose to watch the whole video tutorial at first, then they started to try to replicate the task on the smartphone depending on their short-term memory. When they forgot several steps, they came to the video tutorial to figure it out. They appreciated that videos could \textbf{provide the overview of the entire process}. The others chose to finish the task with the video tutorial playing at the same time. The difference in the choices may contribute to their memory level and habits.} 
Compared to not watching the video, participants reported that watching the video tutorial is more intuitive and makes them feel more confident to some extent than trying by themselves without any support. 

However, participants mentioned that it was still not as good as being taught by their children. \rv{The major complaint was the lack of control over the speed the video plays. They had to go back and forth in order to catch up with the video tutorial. While, this challenge would be in the interactive tutorial expressed by participants, which is synchronous with their pace and easier to follow and learn.}

The tutorial does not highlight the components where the interaction happens. Since we assume that in daily life, help-givers would simply provide screen recording videos without any annotations. Therefore, it leaves the user to discover where to click on. \rv{Besides, the speed of the screen recording made by help-givers may probably be too fast for older adults since it is recorded according to their usage habits.} ``\textit{Sometimes it is so quick, maybe just because we older adults are not familiar with smartphones and have slower reaction speed than the young.}''- P10.

\subsubsection{\rv{Features} of Synapse (Basic)}
Participants highly praised Synapse (Basic). They considered that this mode could help them achieve their goal \textbf{fast}. ``\textit{It could save me a lot of time. I could directly know where to click on and save many procedures, especially when I am busy with caring for my grandchildren.}''- P6. \rv{P6 has retired but offered to help his son to take care of the young grandchildren. It happens a lot in Chinese culture. It is important to understand that this group of older adults may have a specific need of saving time.} Besides, they felt it was convenient and simple to use this mode to complete the task since the highlight feature is clear and intuitive, and easy to understand without extra mind burden. ``\textit{I have \rv{low vision}, this mode really helps me find the exact icon, I used to get lost in these colorful icons.}''- P7. ``\textit{I think that my thinking ability is getting slower and slower now, but this approach is clear, helps me learn faster, and make the process simpler.}''- P5. Moreover, it could lower the barrier of learning \rv{ and be extremely user-friendly for those who are not familiar with smartphones.} ``\textit{Even if I am not familiar with the app and not good at using smartphones, I could still learn easily.}''- P2. \rv{``\textit{If one who is not familiar with the smartphones, in Trial-and-Error mode, he will keep saying "could not find it", which is not good as in the Basic mode.}" - P15} In the meantime, for the feature of protecting them from the wrong click, participants appreciate the \textbf{correctness} and feel comfortable with it. `\textit{Sometimes I am afraid of clicking wrongly on smartphones by myself. This is great, it is always correct and makes me feel comfortable.}''- P9. Participants also mention the usefulness of voice instruction because they are able to know more information to help them understand the current operation and feel more confident when they hear the voice instruction consistent with the visual one. More importantly, participants felt that this approach could \textbf{motivate them to learn} and help them to \textbf{memorize} the procedure to some extent. ``\textit{This makes the learning process not so difficult, so I feel more willing to learn.}''- P15. ``\textit{Some older adults like me, we have poor memory, when we always forget, we will feel impatient and become annoyed, then we did not want to learn it anymore. But with this approach, such a step-by-step way is quite clear, and give me more time to react, it would not move on until I react, I feel that I can master more as time goes by.}''- P5. ``\textit{I believe that after a few more times, I can memorize how to do it and operate it by myself.}''- P13. 

Participants reported the limitation of this approach including the \rv{restrictions on clickable areas}, limited performance of visualization, and \rv{fear of losing the opportunity to exercise the brain}. During the trial using the Basic mode, P2 naturally asked the questions that when she would like to try clicking other areas, she felt restricted with this mode. ``\textit{What if I do not want to click on this area, and want to try it by myself.}''- P12. Although we allow users to terminate the current running tutorial at any time, once they terminate it, the following support disappears as well. As for the visualization format, they expressed that it would be better if it is more visible since the current red rectangle' salience decreases when the icon's color is trending to be red. Besides, participants were concerned that this convenience may not benefit their cognitive ability. ``\textit{Anyway, the only bad thing is that you don't have to use your brain.}''- P4.

\subsubsection{\rv{Features} of Synapse (Trial-and-Error)}
 Participants reported that this mode is \textbf{inspiring} and faster than searching information online for solutions.
 As for the feature of providing Trial-and-Error support with voice assistance, participants appreciated its conformation to natural usage habits, convenient, quick feedback, and felt less burden to ask for others' help. ``\textit{When I really could not find where to click on, it helps me a lot. I love the way that providing support when in need.}''- P1. ``\textit{I could just get instant feedback from it, and I do not need to bother other people to help me. I felt not good if I trouble others.}''- P8. ``\textit{I used to ask others to help me, but they are not always available.}''- P12. 
 As for the feature of providing feedback and error recovery approach, participants reported that they feel more confident and affirmed to have the support during the entire process. ``\textit{I like trying by myself, and I feel confident that it could give you feedback when you click wrongly and give you hint when you get lost.}''- P1 ``\textit{When I clicked on the right place, I feel happy and confident seeing the prompt with a beep. It is the feeling that being encouraged and affirmed. And if you made a mistake, you also get an alert. I would feel a tightening of my heart and I'm glad to have this to remind me before leading to any severe consequence.}''- P18.
 As for the feature of conducting "back" operations at any time through the voice assistant, participants considered it useful and effort-saving. ``\textit{The function of "back to the previous step" helps quite a lot. By just saying "back", it would turn to the previous page and if it is still not on the right page, I could continue to go back until it shows the confirmation prompt of the correct prompt.}" - P14 
 
 Participants further reported that it would \textbf{encourage them to learn and memorize} since they keep thinking during the process instead of just following the tutorial. ``\textit{I put my heart when I try to complete the task, so it impresses me. Next time, when I try it by myself, I could remember it clearly and complete it more proficiently.}''- P17. Participants also felt \textbf{encouraged and motivated} by the novelty and interactive support of the Trial-and-Error mode. ``\textit{I felt very willing to learn, very willing to learn. I never heard of or tried this tool before, it was too interesting, and it really motivates me to try more tasks using this mode.}" - P7 ``\textit{I felt it was vivid and not boring.}" - P5. \rv{Besides, the \textbf{flexibility} of the Trial-and-Error mode was also echoed by participants. ``\textit{I could make some modifications by myself through this approach. Especially for online shopping, I could add different goods to my shopping cart.} - P15.}
 
 Moreover, they appreciated this approach because of their spirit of "not giving in to old age" and would like to learn more. ``\textit{I feel that I am still willing to learn. Currently, I feel that I am not that old. The basic mode is good, but I want to learn more.}" - P4 Participants also commented highly on Synapse (Trial-and-Error) as the future development trend.

The drawback of this mode reported by participants includes the limited performance of visualization as the one in basic mode and feeling not fast enough when they are in urgent need. ``\textit{It still appears to be slow if I am in a hurry, I may give up the Trial-and-Error approach in that circumstance.}" - P1

\subsubsection{Errors Participants Made \rv{When Completing Tasks without Any Support}}

\begin{figure*}[htb!]
  \centering
  \includegraphics[width=\linewidth]{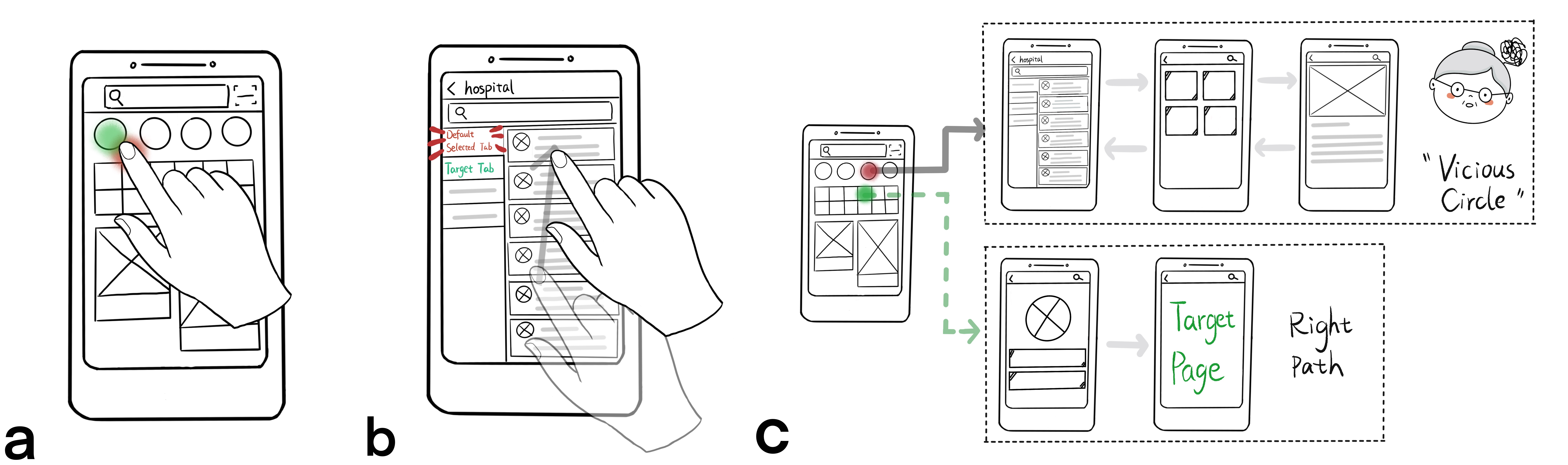}
  \caption{Errors participants faced \rv{when conducting Trial-and-Error without any support}. (a) shows participants often click outside the target icon, (b) shows participants are hard to find the target icon if the target tab is not the default one. Participants would keep rolling up and down in the default domain, (c) shows participants could not return to the right page where they make mistake. They back to the wrong page and keep trying to start from it, which gets them into a  wrong circle.}
  \label{fig:error}
\end{figure*}

\textbf{Participants may not precisely click on the right icon.} The size of icons varies. Participants often clicked outside the icon's range if the icon is small or in an irregular shape as Fig.~\ref{fig:error} a shows. Especially, participants typically did not know the mechanism of the triggering operation in applications. They felt annoyed and puzzled when their click did not take effect. After clicking, they waited for several seconds and if it did not have an action, they muttered "why did not it react?", and clicked again. When they experienced this several times, they became impatient and did "crazy clicking", which means that they non-stop fast clicked.  It did work sometimes, since during the "crazy clicking", there is a chance that they click on the right place which could trigger the following action. However, it also has the risk that the application may recognize multiple clicks at the same time, and execute several clicks at one time. Participants show more worried and panic about this unexpected consequence and keep asking questions like "what happened? why am I on this page?". Participants could solve the problems of \rv{clicking but turning no reaction} after several trials, but they had difficulty solving the severe consequence after the "crazy clicking". They had to totally quit the application and redo it from the beginning.

Another circumstance is that icons were placed so closely together that participants \rv{with low vision} often clicked on the adjacent icons of the intended targets. In addition, when they were confused about the textual descriptions of the icons, they may choose the wrong one. 

\textbf{\rv{It is hard for participants to find the targeted icon in another tab.}} 
Due to the unfamiliarity with the design and usage of smartphones as well as the poor understanding of the selecting logic in apps, they often make mistakes when there is a need to switch tabs. As Fig.~\ref{fig:error} b shows, their target icon is actually within the target tab which is below the default selected tab. But participants kept rolling up and down with the default selected tab. 

\textbf{Participants get lost when conducting return action.} 
During the regular procedure of using an application, return is a common action either for recovering an error or coming back to the previous pages for selecting another option. We observed that participants typically stop on the wrong page and keep trying starting from this wrong page. Therefore, they would hardly get away with a "vicious circle" as Fig.~\ref{fig:error} c shows. 

\section{Discussion}

\subsection{Key Takeaways}
Compared to using the video tutorials, older adults completed tasks faster with  Synapse (Basic) (significant) or Synapse (Trial-and-Error) (trending significantly) based on time spent in session 1. This suggests that both interactive guidance and trial-and-error support could save time for older adults in completing tasks on smartphones. \rv{Just on saving time, the remote assistance app APPAmigo could also achieve the same goal by completing tasks automatically with the approach of recreating the events directly from the help giver's device~\cite{Vyas2019AppAmigo}. Although it may not motivate older adults to learn, it would be useful when an emergency situation occurs and help them solve the problem quickly. Future work may consider incorporating the automation process to better fulfill older adults' need for saving time.}

While in session 2, our results show that older adults completed the tasks fastest, though not significantly, on the app learned with Synapse (Trial-and-Error) in session 1. 
This finding echos the finding of a previous study that compared a video and an interactive tutorial in teaching older adults to use smartphones and found no significant differences~\cite{ribeiro2014efficiency}. Possible reasons might be that even with interactive guidance and trial-and-error support in session 1, older adults still only had limited time to use the apps with this support. Thus, the benefits of the support might not be reflected in session 2.
More research is warranted to investigate the long-term learning effects of the proposed interactive and trial-and-error support. 

In session 2, while older adults made the fewest mistakes with the app they learned with Synapse (Trial-and-Error) in session 1, we did not find a significant difference in errors between video and Synapse (Basic). One possible reason could be that the features designed in Synapse (Basic) such as precise click calibration made older adults easier and faster to complete the task in session 1 (as shown in Fig.4). They probably did not have time to learn different apps as they did in both video and Synapse (Trial-and-Error) conditions. This finding suggests the importance of supporting trial-and-error in addition to merely providing interactive guidance if the end goal is to help old adults complete not just the same task they receive interactive guidance for but also other similar tasks with the apps. 

We identified the following features that might have helped older adults in completing tasks. \rv{In the Video approach, the containing voice instructions helped more than the visual information in the screen recording. Although participants appreciated that it provided an overview of the entire process, it still relied on participants to match the content of the video tutorial with the real application they are using by themselves, similar to the previous approaches~\cite{Pongnumkul2011, Chi2012Mix}. The problem may become more challenging when there was inadequate visual salience, 
such as when some UI components went unnoticed on a page~\cite{yu2020maps}.} In Synapse (Basic), \rv{audio instructions in everyday language created by help-givers were shown along with the highlighted visual cues} to older adults with additional information. Such in-app interactive multimodal instructions were not investigated in prior work, such as EverTutor~\cite{Wang2014EverTutor}. Older adults commented that they felt reassured when the voice instructions and visual clues were consistent as they were relieved from fear and concerns caused by the lack of knowledge and confidence~\cite{vaportzis2017older}. Moreover, we provided support for older adults to click more precisely \rv{in the Basic Mode} by applying the automate calibration algorithm, since physical declines including vision impairments and mild motor impairments (e.g., difficulty in finger movement) make it hard for older adults to touch on targets with precision~\cite{Mohadis2014ASUB, vaportzis2017older, Pang2015OAADS}. Indeed, with the support, older adults felt they could finish the entire task more easily and smoothly without fewer mistakes or extra burdens. Furthermore, we designed a robust enhancement mechanism to check the consistency between the current click position and the recorded tutorial \rv{in case issues, such as unavailable markup~\cite{Sereshkeh2020VASTA}, occur to improve the stability.} \rv{It is worth noting that, although the Basic mode provided interactive guidance, it also prevented users from free exploration and making any mistakes. Some participants were concerned about losing the opportunity to exercise their brains, which was important to many older adults as found in prior work~\cite{Broady2010ComparisonOO}.}
\rv{In Synapse (Trial-and-Error), older adults felt they completed tasks faster and easier. They also felt more enjoyable and confident and were more willing to continue to learn than the Video approach although the performance of Synapse (Trial-and-Error) and the Video approach was not significantly different. Participants preferred voice interfaces to other approaches, such as clicking and touching the screen~\cite{Brewer2016Exploring, Constantin2019Why, Hosseinpanah2018Empathy, Kowalski2019Older, Schl2013Exploring}. In addition to the natural language interaction (e.g., voice assistance), older adults highly praised the error-recovery feature, which encouraged them to perform trial-and-error that they would have otherwise apprehended. When older adults were unable to find where to click, the instant highlight triggered by natural conversations (e.g., ``can't find'') showed them the exact position to click. Error-recovery support such as back function allowed them to recover to the last correct position, where they could safely proceed. While previous work found that older adults are hesitant to conduct trial-and-error due to factors such as lack of encouragement for self-diagnosis~\cite{fan2018guidelines} and self-perception of changes in abilities~\cite{Franz2019Perception}, our findings provided encouraging evidence that older adults are more willing and confident to use trial-and-error approach if they are well supported with interactive error-recovery mechanism, such as the ones explored in our work, and were assured that errors would not result in severe consequences~\cite{barnard2013learning}.}

\subsection{Design Considerations}

\subsubsection{Interactive Guidance} 
\textbf{Better visual indicators}. We adopted a color-based visual highlight (e.g., a red rectangle) to annotate the place where older adults should click. According to older adults' feedback from the interviews, the current visual performance may be limited in the following circumstances. The first one is that the color-based highlight approach is ineffective when the background color is similar to the highlight color. We could address this problem by applying computer vision technology to dynamically change the color of highlighting components in contrast with the main color of the current page. The second one is that the appearance of the current highlight might be too brief for older adults to catch. Participants reported that if they did not grasp the moment of appearance, they may perceive the highlight as part of the interface. One possible improvement suggested by participants was to have a dynamic zooming effect that gives them more time to notice the highlight one. \rv{Moreover, animating visual indicators such as blinking shows in ~\cite{Wang2014EverTutor} and putting a semi-transparent layer except for the part that needs to be pressed may also help highlight where to click. }

\textbf{More flexible mode switch}. Older adults worry about missing the opportunity of \rv{exercising} their brains during the use of basic mode while feeling troublesome to trigger for most of the steps using Trial-and-Error mode if they are not quite familiar with the task. The current design only enables older adults to choose one mode in the first place. \rv{The expectation expressed by older adults aligns with the previous research, that older adults prefer a more independent and flexible learning approach these days~\cite{Pang2021Association}.} We may provide a more flexible mode switch method to make older adults change between these two modes at any time during their usage. 

\textbf{Better mental model of the overview of the entire process}. Older adults appreciated the feature of the video approach that it could provide an overview of the entire process. \rv{Indeed, previous work MixT, which combined the static and video clips together to provide more effective tutorials is appreciated that it helps users to avoid  missed steps~\cite{Chi2012Mix}}. While users still need to match the content of the tutorial with the real application by themselves. One possible way to incorporate the feature into the interactive tutorial is having a floating pages list at the top of the page, showing the position where the current tutorial suites in the entire process.

\rv{\textbf{Personalization}. We found that older adults with different backgrounds had different needs. For example, some of them who lived in nursing homes or lived alone tended to want to learn to use smartphones to keep their brains sharp. While some of them who lived with their children and helped babysit their grandchildren focused more on the efficiency of completing tasks on smartphones to save time. Thus, future work should investigate ways to better support diverse and personalized needs of smartphone use among older adults, which was also suggested as one of the approach to help older adults adopt mobile banking~\cite{inbook}. Previous work has investigated personalization in Ambient Assisted Living solutions as well, ranging from User Interface modifications, to sending messages, to the possibility of changing the state of appliances and devices available in the surrounding context.~\cite{Chesta2017Enabling}. Besides the above personalizable features, inspired by the previous remote assistance tool AppAmigo~\cite{Vyas2019AppAmigo}, the automatic completion of a task could be supported if older adults press for time. Moreover, the real-time help feature which could be achieved by live screen sharing applications \rv{(e.g., InkWire~\cite{InkWire}, TeamViewer~\cite{TeamViewer}, or AirDroid~\cite{AirDroid})} should also be considered as one of the aspects to be incorporated to prevent the emergency situation.}

\subsubsection{Trial-and-Error}

\textbf{Error recovery support}.
Our study found that older adults felt more confident and more willing to do trial-and-error with the support provided by Synapse (Trial-and-Error), which confirms the idea that older adults may be more willing to try if they are supported to be more confident and errors would not lead to severe consequences~\cite{barnard2013learning}. Our current error recovery support enables older adults to go back to the previous page by using natural conversation (e.g., by saying ``back''). They were able to repeat this voice command until they were back to the right page, they would receive visual and audio prompts so that they would know the right page to stop. However, if they deviated too far away from the correct step (e.g., tried too many steps), going back one step at a time may be inefficient. One alternative approach could be that when the user wants to return to the right page, the service could help the user resume to the right page directly. However, more research is needed to understand which one older adult would prefer.

\rv{Furthermore, we found that older adults often entered into the wrong tab as Fig.~\ref{fig:error} b shows or returned to the wrong page as Fig.~\ref{fig:error} c shows and continued to struggle with the wrong start. This was probably due to low information scent similar to what Yu and Chattopadhyay discovered when studying how older adults use map apps~\cite{yu2020maps}. Future work should consider adding a reminder feature to detect the repetitive wrong operation with a wrong start to better support the trial-and-error approach.}

\textbf{Further careful and detailed guidance}.
According to our observation, older adults got impatient when things did not go well as they suggested. For example, when they found that the icon did not react to their clicks, sometimes they did "crazy clicking". We may include more careful and detailed guidance for further support. For example, we could give direction information such as "move left slightly" for help. This concrete information may help them know more about how to calibrate their operation. During the trial-and-error approach, participants also mentioned that they would like a guide for the next step. For example, we could circle a range and indicate that the next step is \rv{within the area} to narrow down the scope of trial-and-error.

\rv{
\subsection{Approaches for Older Adults to Discover, Access, and Use Synapse in Their Real World Scenarios}
\label{sec:realworld}
In this work, we focused on understanding older adults' experiences when using both interactive guidance and trial-and-error support compared to the video tutorial. To better answer our research question, we simulated how older adults would request for help from help-givers via Synapse and how help-givers would send the tutorial to older adults via Synapse, as shown in Figure \ref{fig:systemUsage}. Specifically, in the user study, the interactive tutorials were created using Synapse by the researchers (i.e., the help-givers) and were shown to older adults (i.e., the help-receivers) on the same smartphone. 

In real world scenarios, Synapse will be made available through app stores (e.g., Google Play Store) as shown in Figure \ref{fig:systemUsage}. Older adults who have experience with app stores could download it from the app stores. Alternatively, help-givers, such as older adults' children and friends, can help them download and install Synapse in the first place. If being available, they can help older adults in person and make sure that older adults are able to use the Synapse app independently. If not, alternatively, they can send a download link to older adults, and instruct with the following procedures via phone calls. In addition, we should design an initial interactive tutorial for the novice to have a quick look at how Synapse works in future work. Help givers will also install Synapse on their phones. For older adults who need help, they can use Synapse to send a request via chat functions. Upon receiving the request, help-givers can demonstrate how to use an app with Synapse running in the background and capturing all user interactions and voice instructions. Help-givers can then send the interactive guidance through their Synapse to the older adults (i.e., help-receivers), who can then choose to use either Basic Mode or Trial-and-Error Mode. Future work should investigate how best to enable the interaction between help-receivers and help-givers, for example, how to allow help-givers to ask questions while using the interactive guidance. 

As older adults may need help with learning to use the same app, future work should may also consider creating a community to save the efforts of creating interactive guidance by allowing help-givers to share their interactive guidance to more older adults and allowing older adults to search and use interactive guidance from the community. Older adults may also provide their comments and ratings to help filter out low quality tutorials. 

}
\subsection{Limitations and Future Work}
\label{sec:limt}

Our work is the first to provide both interactive guidance and trial-and-error support based on older adults' needs to help them complete tasks on smartphones. While our results show the promise of the related features in supporting older adults, the current work is limited in the following aspects. The current remote mode requires that the screen layout between help-giver and help-receiver to be consistent including having the same app version. Future work can combine the computer vision technology employed in EverTutor~\cite{Wang2014EverTutor} to overcome potential challenges introduced by differences in apps and OS versions. As a first step to investigate the benefits of interactive guidance with trial-and-error support for older adults, our current implementation \rv{only tests the use of the most common gesture - single tap gesture during our study, but the logic behind the implementation is similar as we discussed in Sec. ~\ref{sec:help-giver}.} Future work should consider how to better capture more complex gestures and present corresponding visual indicators. \rv{The error recovery mode would not work for interaction beyond the front-end part of the app (e.g., it could not "undo" a doctor's appointment or cancel the food order that was sent to a server). Further work can consider providing notifications before executing irreversible operations. Our current robustness enhancement design (Sec.~\ref{sec:helpreceiver}) was shown to help reduce unexpected errors. However, this approach requires older adults to understand what happened when the prompt occurs and react to it properly, which might add an extra burden on them.} 

\camera{We acknowledge that the limited samples in our user study were unlikely to represent all older adults, whose backgrounds can be more diverse. Firstly, while we recruited participants from different sources, the majority of them were recruited from social media platforms and few were from nursing homes. Older adults in nursing homes usually have caregivers to take care of their daily lives and tend not to have a strong motivation to learn to use smartphones independently. Secondly, the majority of our participants were in good health and did not have cognitive or perceptual declines that prevented them from completing our studies independently. In contrast, older adults with varied impairments might adopt different strategies when using smartphones and have different preferences than our participants. Moreover, older adults' occupations before retirement might also affect their mental models when using smartphones. As a result, future work should investigate how older adults with more diverse backgrounds use and perceive the two types of interactive guidance (i.e., Basic and Trial-and-Error modes) that our work introduced through the Synapse app to understand how to design for older adults with dynamic diversity \cite{Gregor2002Designing} and cater to older adults' needs beyond accessibility~\cite{Knowles2021TheHarm}.}

\rv{Our current study focused on older adults (i.e., help-receivers) and did not evaluate whether help-givers could create interactive tutorials with our app. However, it is relatively straightforward for help-givers to demonstrate how to use an app, which is similar to what they would typically use it. Nonetheless, because we did not involve help-givers in our user study, we had limited insights into the interaction processes between the help-givers and older adults. Designing smooth interactions between help-givers and help-receivers would require a substantial user need identification from both sides. As shown in~\cite{Kleinberger2019Supporting}, help-givers and help-receivers may have different preferences. Moreover, help-givers might not be always available. Thus, future work should also investigate ways to support  asynchronous collaboration that allow help-givers to provide assistance and help-receivers to receive assistance asynchronously. Moreover, more scenarios of how Synapse can be accessed and deployed outside the research context should be investigated further.} 
Furthermore, the interactions between help-givers and help-receivers were limited to the demonstrations and there was no real-time communication support between them. However, older adults may still need further explanations from help-givers when they still have difficulty with interactive guidance and trial-and-error support. Thus, future work should consider how to cater to such real-time interaction needs between help-givers and help-receivers (e.g., video or audio communication support). 

Lastly, the current user study was relatively short and had limited ability to understand the learning effects of the novel features that we introduced the Synapse app. Toward this goal, future work can deploy the two modes of Synapse (i.e., Basic and Trial-and-Error) into older adults' daily lives and conduct a longitudinal study.

\section{Conclusion}
To better support and motivate older adults to learn smartphones, we designed an app-agnostic service, \textit{Synapse}, which allows a help-giver to create an interactive tutorial with voice instructions by demonstration and the help-receiver to complete the same task following the support. We designed several features including precise click calibration, robust enhancement, trial-and-error support, and error recovery support in the two modes Synapse (Basic mode) and Synapse (Trial-and-Error mode). We then conducted a controlled experiment with 18 older adults who are 60 and over. Our quantitative and qualitative
results show that Synapse provided better support than the traditional video approach and enabled participants to feel more confident as well as motivated. Our results also show that the majority of participants preferred Synapse (Trial-and-Error mode) due to its conformation to natural usage habits with voice assistance, visual and audio feedback, and promotion to learn, indicating that with sufficient support, older adults actually enjoy trial-and-error. Moreover, participants who preferred Synapse (Basic mode) affirmed the usefulness of voice instruction during the interactive tutorial, the benefit of time-saving and correctness as well as the low learning barrier.

\bibliographystyle{ACM-Reference-Format}
\bibliography{main.bib}
\end{document}